\documentclass[doublecol]{epl2}

\title{Reply to the Comment by A.J. Leggett and Anupam Garg}

\shorttitle{Reply to the Comment by A.J. Leggett and Anupam Garg}

\institute{ \inst{1} Beckman Institute, Department of Electrical and Computer Engineering and Department of Physics\\ University
  of Illinois, Urbana, Il 61801\\ \inst{2} Institute for Advanced Simulation, J\"ulich Supercomputing Centre, Research Centre
  J\"ulich\\ D-52425 J\"ulich, Germany, EU\\ \inst{3} Department of Applied Physics, Zernike Institute for Advanced Materials,
  University of Groningen\\ Nijenborgh 4, NL-9747 AG Groningen, The Netherlands, EU }

\author{K. Hess\inst{1}\thanks{E-mail: k-hess@illinois.edu} \and K. Michielsen\inst{2}\thanks{E-mail:
 k.michielsen@fz-juelich.de} \and H. De Raedt\inst{3}\thanks{E-mail: h.a.de.raedt@rug.nl} } \shortauthor{K. Hess, K. Michielsen
 and H. De Raedt}

\received{17 May 2010}
\acceptedinfinalform{5 August 2010}
\onlinepub{3 September 2010}


\abstract{}%

\begin{document}

\maketitle

In their Comment~\cite{LEGG10} on our Letter~\cite{HESS09}, Leggett and Garg claim that they have introduced in their original
paper (LG1) a dependence on measurement times. They also claim that Eqs.(HMDR1) and (LG2a) can therefore not be linked in such a
way that the arguments of \cite{HESS09} can be transcribed. However, (LG1) distinguishes only three time differences, and all
experimental results corresponding to the same time differences are identically labeled and therefore treated as mathematically
identical. We therefore can not agree with the argumentation of Leggett and Garg: except for a change of nomenclature
Eqs.(HMDR1) and (LG2a) are the same. A more extensive discussion of this point can be found in Ref.~\cite{RAED09a}.

According to Boole~\cite{BO1862}, it is logically and mathematically impossible to violate the inequality
\begin{equation}
|F_{12}\pm F_{13}|\leq1\pm F_{23}, \label{Boole} \end{equation}
where $F_{ij}=\sum_{\alpha = 1}^M A_{i,\alpha}A_{j,\alpha}/M$,
$1\leq i<j\leq 3$, $A_{i,\alpha}=\pm 1$ and $1\leq\alpha\leq M$ labels the different experimental runs that collect triples of
data $(A_{1,\alpha}$, $A_{2,\alpha}$, $A_{3,\alpha})$~\cite{BO1862}. The Bell and the Leggett and Garg inequality have the same
structure as Eq.~(\ref{Boole}), with $i=1,2,3$ labeling the polarization settings or the three time differences, respectively,
and they are contained in the work of Boole~\cite{BO1862}.

Boole proved the following. If a one-to-one correspondence can be established between the binary variables $(A_{1,\alpha}$,
$A_{2,\alpha}$, $A_{3,\alpha})$ of the mathematical description and each triple of binary data collected in each experimental
run $\alpha$, then it is impossible to violate Eq.~(\ref{Boole}). However, in EPR-type or Leggett-Garg-type laboratory
experiments, the data are not collected in triples but in pairs $(A_{1,\alpha}^{(1)},A_{2,\alpha}^{(1)})$,
$(A_{1,\alpha}^{(2)},A_{3,\alpha}^{(2)})$, $(A_{2,\alpha}^{(3)},A_{3,\alpha}^{(3)})$ with corresponding averages $F_{12}^{(1)}$,
$F_{13}^{(2)}$ and $F_{23}^{(3)}$, respectively, and without any further assumptions, $F_{12}^{(1)}$, $F_{13}^{(2)}$ and
$F_{23}^{(3)}$ are not constrained by Eq.~(\ref{Boole})~\cite{RAED09a}.

Leggett and Garg do not refer to Boole's one-to-one correspondence nor to any other point of Boole's work, but instead introduce
a variety of other conditions and phrases (such as macro-realism and induction) by which they justify their use of the
inequality Eq.~(\ref{Boole}). These conditions would have to give a physical justification for dropping the superscripts $(1)$,
$(2)$, and $(3)$, thereby re-establishing the one-to-one correspondence that is required for Eq.~(\ref{Boole}) to hold.
Inequalities of the type Eq.~(\ref{Boole}) in combination with the postulated conditions are then believed to be useful for
testing whether these conditions are satisfied in a laboratory experiment.

Key to the reasoning of Leggett and Garg is their statement $S =: $``the properties of physical ensembles depend only on their
preparation''~\cite{LEGG10}.  These properties are described in their and in our paper by binary data sets. Assuming the
statement $S$ to be true is then sufficient to guarantee a one-to-one correspondence between triples of binary variables
$(A_{1,\alpha},A_{2,\alpha},A_{3,\alpha})$ and the experimental data. Hence, according to Boole~\cite{BO1862}, it is logically
and arithmetically impossible to violate inequalities such as those of Bell and Leggett and Garg. On this point we agree with
Leggett and Garg.

However, there exists no reason, logical or physical, that statement $S$ always needs to be satisfied. $S$ represents an
assumption and restricts the discussion to situations in which the actual act of measurement cannot affect the outcome. In our
opinion, assumption $S$ may indeed be applied to the elements of reality emanating from the source and often denoted by
$\lambda$, but it is a giant leap to apply $S$ to the binary data that are found by involvement of sophisticated measurement
equipment and that may therefore be significantly influenced by that equipment. For example, that equipment may contain some
form of ``clock" that impresses the measurement time $t_m$ on the outcomes of the measurements so that we are justified to
introduce a time label and then deal with functions $A_{i,\alpha,t_m}$, as is customary for stochastic processes~\cite{Willi}.

The example with doctors and patients presented in our Letter\footnote{the word ``each'' in the second line of page 5 should be
read as {\bf ``, each doctor''}}, based on every-day experience and requiring knowledge of elementary arithmetic only, is
macro-realistic in the common sense of the word but not if macro-realistic is used in the narrow sense of the statement $S$
because of the local influences on the data.
The example shows that a violation of a Boole inequality can be accomplished by
dropping Boole's {\bf particular condition of the one-to-one correspondence}.
Other explicit examples pertaining to EPR experiments are given in Ref.~\cite{RAED09a}.
A rigorous treatment of the problem within the framework of Kolmogorov's probability theory
can be found in Ref.~\cite{KHRE}

Note that our properly prepared ``physical ensemble'' of doctors and patients is
definitely realizable, which proves that the statement $S$ of Leggett and Garg does not always need to be satisfied, even in the
macroscopic realm.

\end{document}